
\magnification=\magstep1
\parskip=6pt
\input amstex
\documentstyle{amsppt}
\NoBlackBoxes

\def\grad{\operatorname{grad}}
\def\leaderfill{\leaders\hbox to 1em{\hss.\hss}\hfill}

\document
\null
\vskip 2truein
\centerline{\bf VECTOR FIELDS AND THE UNITY OF MATHEMATICS AND PHYSICS}
\medskip
\centerline{by}
\bigskip
\centerline{\bf Daniel H.\ Gottlieb }
\bigskip
\centerline{\bf Department of Mathematics}
\bigskip
\centerline{\bf Purdue University}
\centerline{\bf  West Lafayette, Indiana}
\vskip .25truein
\noindent
{\bf Abstract}

We give an argument that magnetic monopoles should not exist. It is
based on the concept of the index of a vector field. The thrust of the
argument is that indices of vector fields are invariants of space-time
orientation and of coordinate changes, and thus physical vector fields
should preserve indices. The index is defined inductively by means of an
equation called the Law of Vector Fields. We give extended
philosophical
arguments that this Law of Vector Fields should play an important role
in mathematics, and we back up this contention by using it in a
mechanical way to greatly generalize the Gauss--Bonnet theorem and the
Brouwer fixed point theorem and get new proofs of many other theorems.
We also give some other suggestions for using the Law and index in physics.

\vfill\eject
\baselineskip=18pt
\noindent
{\bf 1.\ \ Introduction}

We will give an argument that magnetic monopoles do not exist.
Magnetic monopoles were predicted by Dirac {\cite F}
based on an alteration of
Maxwell's equations which made them more symmetric.
Despite Dirac's ideas, magnetic monopoles have not been found in nature.
This is the case even though Dirac used similar considerations to predict
antiparticles and other phenomena.

Our argument goes as follows:\ \ First we state a general principle which we
will assume holds for every ``physical vector field.''

\proclaim{Principle of Invariance of Index}
The index of any ``physical'' vector
field is invariant under changes of coordinates and orientation of
space--time. It must be physically significant. If the index is
undefined, it signals either radiation or unrealistic physical
hypotheses.
\endproclaim

\proclaim{Consequence} Every ``physical'' pseudo--vector field has index zero
or the
index is undefined.
\endproclaim

Now the magnetic vector field $\overrightarrow{B}$ is a pseudo--vector field.
That means if we change the orientation of space $\overrightarrow{B}$ changes
to
$-\overrightarrow{B}$.
Now Ind$(-V)=(-1)^n$ Ind$(V)$ where $n$ is the dimension of the
manifold on which $V$ is defined.
Thus Ind$(-\overrightarrow{B})=(-1)^3$ Ind$(\overrightarrow{B})$.
So either Ind$(\overrightarrow{B})$ is not defined or
Ind$(\overrightarrow{B})=0$.

Now a magnetic monopole will give rise to a $\overrightarrow{B}$ with index
$\pm 1$.
As this is inconsistent with the Invariance of Index Principle, we predict
that magnetic monopoles do not exist.

We have some arguments which tend to explain why the principle of invariance of
index is reasonable.
These arguments concern things mathematical rather than physical.
They are not theorems, rather they are predictions and explanations of some
things which will happen or have happened in mathematics.

These considerations lead us to the prediction that a certain equation due
to Marston Morse, {\cite M},
will play a very active role in mathematics, and by
extension physics.
This equation, which we call the Law of Vector Fields was discovered in 1929
and has not played a role at all commensurate with our prediction up until now.

We describe the equation, which we call the Law of Vector fields.  Let
$M$ be a compact manifold with boundary.  Let $V$ be a vector field on
$M$
with no zeros on the boundary. Then consider the open set of the
boundary
of $M$ where $V$ is pointing inward.  Let $\partial_{-}V$ denote the
vector
field defined on this open set on the boundary which is given by
projecting
$V$ tangent to
the boundary.  The Euler characteristic of $M$ is denoted by
$\chi(M)$, and $\Ind(V)$ denotes the index of the vector field.  Then the
Law of Vector Fields is
$$
\Ind(V) + \Ind(\partial_{-}V) = \chi(M),
$$

We propose two methods of applying the law of vector fields to get new results
and we report on their successes.
These successes and the close bond between physics and mathematics encourage
us to predict that the Law of Vector Fields and its attendant concepts must
play a vital role in physics.

\bigskip\noindent
{\bf 2.\ \ The Unity of Mathematics}

We take the following definition of mathematics:

\proclaim{Definition} Mathematics is the study of well--defined concepts.
\endproclaim

Now well--defined concepts are creations of the human mind.
And most of those creations can be quite arbitrary.
There is no limit to the well--defined imagination.
So if one accepts the definition that mathematics is the study of the
well--defined, then how can mathematics have an underlying unity?
Yet it is a fact that many savants see a underlying unity in mathematics,
so the
key question to consider is:

\proclaim{Question}
Why does mathematics appear to have an underlying unity?
\endproclaim

If mathematical unity really exists then it is reasonable to hope that there
are a few basic principles which explain the occurrence of those phenomena
which persuade us to believe that mathematics is indeed unified;
just as the various phenomena of physics
seem to be explained by a few fundamental laws.
If we can discover these principles it would give us great insight into the
development of mathematics and perhaps even insight into physics.

Now what things produce the appearance of an underlying
unity in mathematics?
Mathematics appears to be unified when
a concept, such as the Euler characteristic, appears over and over in
interesting results; or an idea, such as that of a group, is involved in
many different fields and is used in science to predict or make
phenomena precise ; or an equation, like De Moivre's formula
$$
e^{i\theta} = \cos\theta + i\sin\theta
$$
yields numerous interesting relations among important concepts in
several
fields in a mechanical way.

Thus underlying unity comes from the ubiquity of certain concepts and objects,
such as the numbers $\pi$ and $e$ and concepts such as groups and rings,
and invariants such as the Euler characteristic and eigenvalues, which
continually appear in striking relationships and in diverse fields of
mathematics and physics.
We use the word {\it broad} to describe these concepts.

Compare broad concepts with {\it deep} concepts.
The depth of an idea seems to be a function of time.
As our understanding of a field
increases, deep concepts become elementary concepts, deep
theorems are transformed into definitions and so on.
But something broad, like the Euler characteristic, remains broad, or becomes
broader as time goes on.
The relationships a broad concept has with other concepts are forever.

\proclaim{The Function Principle}
Any concept which arises from a simple construction
of functions will appear over and over again throughout mathematics.
\endproclaim

We assert the principle that function is one of
the broadest of all mathematical
concepts, and any concept or theorem derived in a natural way from that of
functions must itself be broad.
We will use this principle to assert that the underlying unity of mathematics
at least partly stems from the breadth of the concept of function.
We will show how the breadth of category and functor and equivalence and $e$
and $\pi$ and de Moivre's formula and groups and rings and Euler Characteristic
all follow from this principle.
We will subject this principle to the rigorous test of a scientific theory:\ \
It must predict new broad concepts.
We make such predictions and report on evidence that the predictions are
correct.

The concept of a function as a mapping $f\colon X\rightarrow Y$ from a source
set $X$ to a target set $Y$ did not develop until the twentieth century.
The modern concept of a function did not even begin to emerge until the middle
ages.
The beginnings of physics should have given a great impetus to the notion of
function, since the measurements of the initial conditions of an experiment
and the final results gives implicitly a function from the initial states of an
experiment to the final outcomes; but historians say that the early physicists
and mathematicians never thought this way.
Soon thereafter calculus was invented.
For many years afterwards functions were thought to be always given by some
algebraic expression.
Slowly the concept of a function of a mapping grew.
Cantor's set theory gave the notion a good impulse but the modern notion was
adopted only in the Twentieth Century.

The careful definition of function is necessary so that the definition of the
composition of two functions can be defined.
Thus $f o g$ is only defined when the target of $g$ is the source of $f$.
This composition is associative:\ \ $(f o g)oh=fo(goh)$ and $f$ composed with
the identity of either the source or the target is $f$ again.
We call a set of functions a {\it category} if it is closed under compositions
and contains the identity functions of all the sources and targets.

Category was first defined by S.\ Eilenberg and S.\ MacLane and was employed
by Eilenberg and N.\ Steenrod in the 1940's to give homology theory its
functorial character.
Category theory became a subject in its own right, it's practitioners joyfully
noting that almost every branch of mathematics could be organized as a
category.
The usual
definition of category is merely an abstraction of functions closed
under composition.
The functions are abstracted into things called morphisms and composition
becomes an operation on sets of morphisms satisfying exactly the same
properties
that functions and composition satisfy.
Most mathematicians think of categories as very abstract things and are
surprised to find they come from such a homely source as functions closed
under composition.

A {\it functor} is a function whose source and domain are categories and
which preserves composition.
That is, if $ F$ is the functor, then $ F(f o g)= F(f)o  F(g)$.
This definition also is abstracted and one says category and functor in the
same breath.

Now consider the question:\ \ What statements can be made about a function $f$
which would make sense in every possible category?
There are basically
only four statements since the only functions known to exist in every
category are the identity functions.
We can say that $f$ is an identity, or that $f$ is a {\it retraction} by which
we mean that there is a function $g$ so that $f o g$ is an identity, or that
$f$ is a {\it cross--section} by which we mean that there is a function $h$
so that $h o f$ is an identity, or finally that $f$ is an {\it equivalence}
by which we mean that $f$ is both a retraction and a cross--section.
In the case of equivalence the function $h$ must equal the function $g$ and it
is called the {\it inverse} of $f$ and it is unique.

Retraction and cross--section induce a partial ordering of the sources and
targets of a category, hereafter called the {\it objects} of the category.
Equivalences induce an equivalence relation on the objects and give us the
means of making precise the notion that two mathematical structures are the
same.

Now consider the self equivalences of some object $X$ in a category of
functions.
Since $X$ is both the source and the target, composition is always defined for
any pair of functions, as are inverses.
Thus we have a {\it group}.
The definition of a group in general is just an abstraction, where the
functions
become undefined elements and composition is the undefined operation which
satisfies the group laws of associativity and existence of identity and
inverse, these laws being the relations that equivalences satisfy.
The notion of functor restricted to a group becomes that of
{\it homomorphism}.
The equivalences in the category of groups and homomorphisms are called
{\it isomorphisms}.

The concept of groups arose in the solution of polynomial equations, with the
first ideas due to Lagrange in the late eighteenth century, continuing through
Abel to Galois.
Felix Klein proposed that geometry should be viewed as arising from groups of
symmetries in 1875.
Poincare proposed that the equations of physics should be invariant under the
correct symmetry groups around 1900.
Since then groups have played an increasingly important role in mathematics
and in physics.
The increasing appearance of this broad concept must have fed the feeling of
the underlying unity of mathematics.
Now we see how naturally it follows from the Function Principle.

If we consider a set of functions $S$ from a fixed object $X$ into a group $G$
we can induce a group structure on $S$ by defining the multiplication of two
functions $f$ and $g$ to be $f*g$ where $f*g(a)=f(a)\cdot g(a)$ where $a$ runs
through all the elements in $X$ and $\cdot$ is the group multiplication in $G$.
This multiplication can be easily shown to satisfy the laws of group
multiplication.
The same idea applied to maps into the Real Numbers or the Complex Numbers
gives rise to addition and multiplication on functions.
These satisfy properties which are abstracted into the concepts of abelian
rings.
If we consider the set of self homomorphisms
of abelian groups and use composition and
addition of functions, we get an important
example of a non--commutative ring.
The natural functors for rings should be ring homomorphisms.
In the case of a ring of functions into the Real or Complex numbers we note
that a ring homomorphism $h$ fixes the constant maps.
If we consider all functions which fix the constants and preserve the
addition, we get a category of functions from rings to rings; that is, these
functions are closed under composition.
We call these functions {\it linear transformations}.
They contain the ring homomorphisms as a subset.
Study the equivalences of this category.
We obtain the concepts of {\it vector spaces} and linear transformations
after the usual abstraction.

Now we consider a category of homomorphisms of abelian groups.
We ask the same question which gave us equivalence and groups, namely:\ \
What statements can be made about a homomorphism $f$ which would make sense
in every possible category of abelian groups?
Now between every possible abelian group there is the trivial homomorphism
$0\colon A\rightarrow B$ which carries all of $A$ onto the identity of $B$.
Also we have for every integer $N$ the homomorphism from $A$ to itself which
adds every element to itself $N$ times, that is multiplication by $N$.

Thus for any homomorphism $h\colon A\rightarrow B$ there are three statements
we can make which would always make sense.
First $N o h$ is the trivial homomorphism 0, second that there is a
homomorphism $\tau\colon B\rightarrow A$ so that $h o \tau$ is multiplication
by $N$, or third that $\tau o h$ is multiplication by $N$.
So we can give to any homomorphism three non--negative integers:\ \ The
{\it exponent}, the {\it cross--section degree}, and the {\it retraction
degree}.
The {\it exponent} is the smallest positive integer such that $N o h$ is the
trivial homomorphism $0$.
If there is no such $N$ then the exponent is zero.
Similarly the {\it cross--section degree} is the smallest positive $N$ such
that there is a $\tau$, called a {\it cross--section transfer}, so that
$h o \tau$ is multiplication by $N$.
Finally the {\it retraction degree} is the smallest positive $N$ such that
there is a $\tau$, called a {\it retraction transfer}, so that $\tau o h$ is
multiplication by $N$.

In accordance with the Function Principle,
we predict that these three numbers will be
seen to be broad concepts.
Their breadth should be less than the breadth of equivalence, retraction and
cross--section because the concepts are valid only for categories of abelian
groups and homomorphisms.
But exponent, cross--section degree and retraction degree can be pulled back
to other categories via any functor from that category to the category of
abelian groups.
So these integers potentially can play a role in many interesting categories.
In fact for the category of topological spaces and continuous maps we can say
that any continuous map $f\colon X\rightarrow Y$ has exponent $N$ or
cross--section degree $N$ or retraction degree $N$ if the induced
homomorphism $f_*\colon H_*(X)\rightarrow H_*(Y)$ on integral homology has
exponent $N$ or cross--section degree $N$ or retraction degree $N$
respectively.

As evidence of the breadth of these concepts we point out that for integral
homology, cross--section transfers already play an important role for fibre
bundles.
There are natural transfers associated with many of the important classical
invariants such as the Euler characteristic and the index of fixed points and
the index of vector fields,\cite{B--G},
and the Lefschetz number and coincidence number and
most recently the intersection number.
And a predicted surprise relationship occurs in the case of cross--section
degree for a map between two spaces.
In the case that the two spaces are closed oriented manifolds of the same
dimension, the cross--section degree is precisely the absolute value of the
classical Brouwer degree.
The retraction degree also is the Brouwer degree for closed manifolds if we
use cohomology as our functor instead of homology, \cite{G1}.

The most common activity in mathematics is solving equations.
There is a natural way to frame an equation in terms of functions.
In an equation we have an expression on the left set equal to an expression
on the right and we want to find the value of the variables for which the two
expressions equal.
We can think of the expressions as being two function $f$ and $g$ from $X$ to
$Y$ and we want to find the elements $x$ of $X$ such that $f(x)=g(x)$.
The solutions are called {\it coincidences}.
Coincidence makes sense in any category and so we would expect the elements
of any existence or uniqueness theorem about coincidences to be very broad
indeed.
But we do not predict the existence of such a theorem.
Nevertheless in topology there is such a theorem.
It is restricted essentially to maps between
closed oriented manifolds of the same dimension.
It asserts that locally defined coincidence indices add up to a globally
defined
coincidence number which is given by the action of $f$ and $g$ on the homology
of $X$.
In fact it is the alternating sum of traces of the composition of the umkehr
map $f_!$, which is defined using Poincare Duality, and $g_*$, the homomorphism
by $g$.
We predict, at least in topology and geometry, more frequent appearances of
both the coincidence number and also the local coincidence index and they
should relate with other concepts.

If we consider self maps of objects, a special coincidence is the fixed point
$f(x)=x$.
{}From the point of view of equations in some algebraic setting, the
coincidence
problem can be converted into a fixed point problem, so we do not lose any
generality in those settings by considering fixed points.
In any event the fixed point problem makes sense for any category.
Now the relevant theorem in topology is the Lefschetz fixed point theorem.
In contrast to the coincidence theorem, the Lefschetz theorem holds essentially
for the wider class of compact spaces.
Similar to the coincidence theorem, the Lefschetz theorem has locally defined
fixed point indices which add up to a globally defined Lefschetz number.
This Lefschetz number is the alternating sum of traces of $f_*$, the
homomorphism induced by $f$ on homology.
This magnificent theorem is easier to apply than the coincidence theorem and so
the Lefschetz number and fixed point index are met more frequently in various
situations than the coincidence number and coincidence indices.

In other fields fixed points lead to very broad concepts and theorems.
A linear operator gives rise to a map on the one dimensional subspaces.
The fixed subspaces are generated by {\it eigenvectors}.
Eigenvectors and their associated eigenvalues play an important role
in mathematics and physics and are to be found in the most surprising places.

Consider the category of $C^\infty$ functions on the Real Line.
The derivative is a function from this category to itself taking any function
$f$ into $f^\prime$.
The derivative practically defines the subject of calculus.
The fixed points of the derivative are multiples of $e^x$.
Thus we would predict that the number $e$ appears very frequently in calculus
and any field where calculus can be employed.
Likewise consider the set of analytic functions of the Complex Numbers.
Again we have the derivative and its fixed point are the multiples of $e^z$.
Now it is possible to relate the function $e^z$ defined on a
complex plane with real valued functions by
$$
e^{(a+ib)}=e^a (\cos(b)+i\sin (b)).
$$
We call this equation de Moivre's formula.
This formula contains an unbelievable amount of information.
Just as our concept of space--time separation is supposed to break down near
a black hole in physics, so does our definition--theorem view of mathematics
break down when considering this formula.
Is it a theorem or a definition?
Is it defined by sin and cos or does it define those two functions?

Up to now the function
principle predicted only that some concepts and objects will
appear frequently in undisclosed relationships with important concepts
throughout mathematics.
However the de Moivre equation gives us methods for discovering the precise
forms of some of the relationships it predicts.
For example, the natural question ``When does $e^z$ restrict to real valued
functions?'' leads to the ``discovery'' of $\pi$.
{}From this we might predict that $\pi$ will appear throughout calculus type
mathematics, but not with the frequency of $e$.
Using the formula in a mechanical way we can take complex roots, prove
trigonometric identities, etc.

There is yet another fixed point question to consider:\ \ What are the fixed
points of the identity map?
This question not only makes sense in every category; it is solved in every
category!
The invariants arising from this question should be even broader than those
from
the fixed point question.
But at first glance they seem to be very uninteresting.
However, if we consider
the fixed point question for functions which are equivalent
to the identity under some suitable equivalence relation in a suitable category
we may find very broad interesting things.
A suitable situation involves the fixed points of maps homotopic to the
identity
in the topological category.
For essentially compact spaces the {\it Euler characteristic} (also called the
Euler--Poincare number) is an invariant of a space whose nonvanishing results
in the existence of a fixed point.
This Euler characteristic is the most remarkable of all mathematical
invariants.
It can be defined in terms simple enough to be understood by a school boy, and
yet it appears in many of the star theorems of topology and geometry.
A restriction of the concept of the Lefschetz number, its occurrence far
exceeds
that of its ``parent'' concept.
First mentioned by Descartes, then used by Euler to prove there were only
five platonic solids, the Euler characteristic slowly proved its importance.
Bonnet showed in the 1840's that the total curvature of a closed surface
equaled a constant times the Euler characteristic.
Poincare gave it its topological invariance by showing it was the alternating
sum of Betti numbers.
In the 1920's Lefschetz showed that it determined the existence of fixed points
of maps homotopic to the identity, thus explaining, according to the
Function Principle,
its remarkable history up to then and predicting, according to our principle,
the astounding frequency of its subsequent appearances in mathematics.

The Euler characteristic is equal to the sum of the local fixed point indices
of the map homotopic to the identity.
We would predict frequent appearances of the local index.
Now on a smooth manifold we consider vector fields and regard them as
representing infinitesimally close maps to the identity.
Then the local fixed point index is the local index of the vector field.
Now we have Morse's formula, which we call the Law of Vector Fields:
$$
\hbox{Ind}(V)+\hbox{Ind}(\partial_{-}V)=\chi (M).
$$
\bigskip\noindent
{\bf 3.\ \ The Law of Vector Fields}

Just as de Moivre's formula
gives us  mechanical
methods which yields precise relationships among broad concepts,
we predict that the Law of Vector Fields will give mechanical
methods which will
yield precise relationships among broad concepts.
The following observation will suggest one method:
The Law of Vector Fields is practically an inductive definition of the
index of
a vector field given the Euler characteristic.
The induction is on the dimension of the underlying manifold.
We say that the index of an empty vector field is zero and the index of a
vector
field on a finite set of points is the number of points.
Then the knowledge of the index on the boundary gives the index in the one
higher dimensional compact manifold.
And the index can be defined on open sets by setting it equal to the index of a
compact manifold with boundary which contains all the zeros of the vector
field.

The method follows:

\item{1.}Choose an interesting vector field $V$ and manifold $M$.

\item{2.}Adjust the vector field if need be to eliminate zeros on the boundary.

\item{3.}Identify the global and local Ind $V$.

\item{4.}Identify the global and local index Ind$(\partial_- V)$.

\item{5.}Substitute 3 and 4 into the Law of Vector Fields.

We predict that this method will succeed because the Law of Vector
Fields is morally the definition of index, so all features of the index
must be derivable from that single equation.  We measure success in the
following descending order: 1. An important famous theorem generalized;
2.
A new proof of an important famous theorem; 3. A new, interesting
result.
We put new proofs before new results because  it may not be apparent at
this time that the new result will famous or important.

In category 1 we already have the extrinsic Gauss-Bonnet theorem of
differential geometry \cite{G3}, the Brouwer fixed point
theorem of topology \cite{G3}, and
Hadwiger's formulas of integral geometry \cite{G3}
,\cite{Had}, \cite{Sp}.  In category 2 we have the
Jordan separation theorem, The Borsuk-Ulam theorem, the Poincare-Hopf
index
theorem of topology; Rouche's theorem and the Gauss-Lucas in complex
variables;
the fundamental
theorem of algebra and the intermediate value theorem of elementary
mathematics; and the not so famous Gottlieb's theorem of group homology,
\cite{G2}.
Of course we have more results in category 3, but it is not so easy to
describe them with a few words.  One snappy new result is the following:
Consider any straight line and smooth surface of genus greater than 1 in
three dimensional Euclidean space.  Then the line must be contained in a
plane which is tangent to the surface, ({\cite{G3}}, theorem 15).

We will discuss the Gauss-Bonnet theorem since that yields results in
all
three categories as well as having the longest history of all the
results
mentioned.  One of the most well-known theorems from ancient times is
the
theorem that the sum of the angles of a triangle equals 180 degrees.
Gauss showed for a triangle whose sides are geodesics on a surface $M$
in
three-space that the sum of the angles equals $\pi + \int_{M}KdM$, where
$K$ is the Gaussian curvature of the surface.  Bonnet pieced these
triangles
together to prove that for a closed surface $M$ the total curvature
$\int_{M}KdM$ equals $2\pi\chi(M)$.  Hopf proved that $\int_{M}KdM$,
where
$M$ is a closed hypersurface in odd dimensional Euclidean space and $K$
is
the product of the principal curvatures must equal the degree of the
Gauss
map $\hat N: M^{2n}\to S^{2n}$ times the volume of the unit sphere.
Then
he proved $2\deg(\hat N) = \chi(M^{2n})$.  (Morris Hirsch in \cite{Hi}
gives credit to Kronicker and Van Dyck for Hopf's result.)
For a history
of
the Gauss-Bonnet theorem see \cite{Gr}, pp. 89-72 or \cite{Sp}, p. 385.

Given a map $f$ between Riemannian manifolds $M$ to $N$ and a vector
field
$V$ on $N$, we can define a vector field on $M$, denoted $f^{*}V$, which
is
the pullback of $V$ by $f$.  For a map $f$ from $M$ into the real line
$R$
the pullback vector field of the unit positive pointing $d/dt$ is just
the
gradient of $f$.

Let $f: M\to R^{n}$ be a smooth map from a compact Riemannian manifold
of
dimension $n$ to $n$-dimensional Euclidean space so that $f$ near the
boundary $\partial M$ is an immersion.  Let $V$ be any vector field on
$R^{n}$ such that $V$ has no zeros on the image of the boundary
$f(\partial
M)$.  Consider the pullback vector field $f^{*}V$ on $M$.  The local
index
on $\partial_{-}V$ turns out to be the local coincidence number of two
different Gauss maps.  Substituting into the Law of Vector Fields
results
in a great generalization of the Gauss-Bonnet Theorem,
({\cite{G3}}, theorem 5).
A very special case is the following, which still is a
generalization
of the Gauss-Bonnet Theorem.  We quote the special case only in order to
avoid introducing notation.  The index of the
gradient of
$xof: M\to R$, where $x$ is the projection of $R^{n}$ onto the $x$-axis,
is equal to the difference between the Euler Characteristic and the
degree
of the Gauss map.  Thus
$$
\Ind(\grad(xof))= \chi(M)-\deg \hat N.
$$
This equation leads to an immediate proof of the Gauss-Bonnet Theorem,
since for odd dimensional $M$ and any vector field $W$, the index
satisfies
$\Ind(-W)= -\Ind(W)$.  Thus the left side of the equation reverses sign
while the right side of the equation remains the same.  Thus $\chi(M)$
equals the degree of the Gauss-map, which is the total curvature over
the
volume of the standard $n-1$ sphere.  Now $2\chi(M) = \chi(\partial M)$,
so
we get Hopf's version of the Gauss-Bonnet theorem.

Note as a by-product we also get $\Ind(\grad(xof)) = 0$ which  is a new
result thus falling into category 3.  Another consequence of the
generalized
Gauss-Bonnet theorem follows when we assume the map $f$ is an immersion.
In
this case the gradient of $xof$ has no zeros, so its index is zero so
the
right hand side in zero and so again $\chi(M)= \deg \hat N$.  This is
Haefliger's theorem \cite{Ha}, a category 2 result.  Please note in
addition that the Law of Vector Fields applied to odd dimensional closed
manifolds combined with the category 2 result $\Ind(-W)= -\Ind
(W)$, implies that the Euler characteristic of such manifolds is zero,
(category 2).  So the Gauss-Bonnet theorem and this result have the same
proof in some strong sense. Also the physical prediction of the
non-existence of magnetic monopoles follows from the same result.

Just as the Gauss-Bonnet theorem followed from pullback vector fields,
the
Brouwer fixed point theorem is generalized by considering the following
vector field.  Suppose $M$ is an $n$-dimensional body in $R^{n}$ and
suppose that $f: M\to R^{n}$ is a continuous map.  Then let the vector
field $V_{f}$ on $M$ be defined by drawing a vector from $m$ to the
point
$f(m)$ in $R^{n}$.  Locally the index of $-V_{f}$  at a zero is a
coincidence index of the Gauss map $\hat N$ and the ``Gauss" map of the
vector field $V_{f}$ given by setting every vector to unit length and
translating to the origin.  If $f$ maps $M$ into itself with no fixed
points on the boundary, then applying the method gives
$$
\Lambda_{f} + \Lambda_{\hat V_f,\hat N} = \chi(M)
$$
This is a category 3 relationship among the main invariants of fixed
point
theory, coincidence theory and the Euler characteristic.  Using the fact
that the coincidence number $\Lambda_{\hat V_{f},\hat N}$ is equal to
$\deg
\hat N - \deg \hat V_{f}$ and the fact from the Gauss-Bonnet theorem
above
that $\deg \hat N= \chi(M)$ we see that the Lefschetz number must be
equal
to $\deg \hat V_{f}$.  If we drop the requirement that $f$ maps $M$ into
itself, we still have $\Ind V_{f} = \deg \hat V_{f}$.  If the map $f$
satisfies the transversal property, that is the line between $m$ on the
boundary of $M$ and $f(m)$ is never tangent to $\partial M$, than $f$
has
a fixed point if $\chi(M)$ is odd (category 1).  This last sentence is
an
enormous generalization of the Brouwer fixed point theorem, yet it
remains
a small example of what can be proved from applying the Law of Vector
Fields
to $V_{f}$.  In fact the Law of Vector Fields applied to $V_{f}$ is the
proper generalization of the Brouwer fixed point theorem.

A second method of producing mathematics from the Law of Vector Fields
involves making precise the statement that the Law defines the index of
vector fields, \cite{G--S}.
In this method we learn from the Law.  The Law teaches us
that there is a generalization of homotopy which is very useful.  This
generalization, which we call {\it otopy}, not only allows the vector
field to
change under time, but also its domain of definition changes under time.
An otopy is what $\partial_{-}V$  undergoes when $V$ is undergoing a
homotopy.  A proper otopy is an otopy which has a compact set of zeros.
The proper otopy classes of vector fields on a connected manifold is in
one
to one correspondence with the integers via the map which takes a vector
field to its index.  This leads to the fact that homotopy classes of
vector
fields on a manifold with a connected boundary where no zeros appear on
the
boundary are in one to one correspondence with the integers.  This is
not
true if the boundary is disconnected.

We find that we do not need to assume that vector fields are continuous.
We can define the index for vector fields which have discontinuities and
which are not defined everywhere.  We need only assume that the set of
``defects" is compact and never appears on the boundary or frontier of
the
sets for which the vector fields are defined.  We
then can define an index for any compact connected component of defects
(subject only to the mild condition that the component is open in the
subspace of defects).  Thus under an otopy it is as if the defects
change
shape with time and collide with other defects, and all the while each
defect has an integer associated with it.  This integer is preserved
under
collisions.  That is the sum of the indices going into a collision
equals
the sum of the indices coming out of a collision, provided no component
``radiates out to infinity", i.e. loses its compactness.

This picture is very suggestive of the way charged particles are
supposed
to interact.  Using the Law of Vector Fields as a guide we have defined
an
index which satisfies a conservation law under collisions.  The main
ideas
behind the construction involve dimension, continuity, and the concept
of
pointing inside.  We suggest that those ideas might lie behind all the
conservation laws of collisions in physics.

\bigskip

{\bf 4.\ \ The Index and Physics}

The success of the concepts of eigenvalue and group and the frequency
of $e$
and $\pi$ in physics suggests the question:\ \ Will the concept of index and
the Law of vector fields play a role in mathematical physics also?
Already in the guise of degree theory the Euler characteristic and fixed points
have appeared in magnetic polarization problems and in defects and textures in
an ordered medium, but we predict that the Law of Vector Fields will be useful
in describing more fundamental things.

We can use the Law to define the index given that we know the Euler
characteristic.
We say the index of an empty vector field is zero.
For zero dimensional manifolds we have a discrete set of points.
For those points for which the vector field is defined, a vector is
necessarily a zero.
If we have a finite number of zeros we define the index to be the number of
zeros, otherwise we say the index is not defined.
Next we look at compact one dimensional manifolds with boundaries.
If the vector field has no singularity on the boundary, we define the index
to be
$$
\hbox{Ind } V=\chi(M)-\hbox{Ind } \partial_{-}V.
$$
Otherwise the index is not defined.
We may extend our definition for $V$ which are not defined or continuous in
the interior of $M$.
We require only that the set of defects, that is the set of zeros, points of
discontinuity of $V$ and points of $M$ where $V$ is not defined, be compact
and disjoint from $\partial M$.
This lack of continuity will allow us to define indices for vector fields
such as an electric field of an electron inside a ball without having to
worry that the field is not defined at the electron.
Next we consider one dimensional manifolds without boundary.
The vector field $V$ is defined on an open set $U$ contained in $M$.
We assume that the set of defects is compact inside of $U$.
So we have a distinction between points in $M$ for which $V$ is not defined.
The points inside $U$ at which $V$ is not defined are called defects.
$V$ is not defined at any point outside of $U$.
We have no name for those points.
Now we can find a compact manifold with boundary which contains the defects
and is itself contained in $U$.
We define the index of $V$ to be the index of $V$ restricted to the manifold
with boundary.
It can be shown that this is well--defined, that is it does not depend on the
choice of the compact manifold which contains the defects, [G--S].

Now we proceed by induction, first to compact two manifolds with boundary and
then to open manifolds with $V$ defined on open subsets $U$,
then for compact 3--manifolds with boundary and then to open 3--manifolds
and so on.
So the index is defined using only continuity, dimension and the concept of
inward pointing.
These topological properties are so basic they are almost physical.
We can think of it this way.
For a body $M$ in space and a vector field $V$ we have the following situation.

\item{a)}Ind $V$ is not defined, in which case there is a defect on the
boundary of $M$.

\item{b)}Ind $V$ is defined, in which case there is no defect on the boundary
of $M$.

\item{c)}Ind $V$ is defined and not equal to zero, in which case there is no
defect on the boundary but there is a defect inside $M$.

If we vary $V$ continuously with time, that is, if we homotopy $V$, we see
that Ind $V$ remains constant if no defects pass through the boundary.
This fact leads to a dynamical picture of defects of vector fields.
We define an index for each connected component of the set of defects by
taking a compact manifold with boundary which contains that component in its
interior and does not contain any other defect and then letting the index of
$V$ restricted to the surrounding compact manifold be the index of the
component.
This is well--defined in the sense that the index does not depend on the
surrounding manifold.
However the index of a connected component of defects may not be defined if
there is no surrounding manifold with boundary which contains no other defect.
This would happen if the component is not compact, or in the rare case that
the component is not isolated from the other defects of $V$.

This picture agrees with the definition of Ind $V$ since Ind $V$ is equal to
the sum of the indices of the components of the defects contained in the
interior of $M$ when every index is defined.
Now under a homotopy these components move around and collide with one another.
There is a conservation law which says that the sum of the indices of the
components going into a collision is equal to the sum of the indices of the
components at the collision is equal to the sum of the indices after the
collision if during the homotopy all the components remain compact and there
are
only a finite number of them.
Thus the index remains conserved unless some component ``radiates out to
infinity.''
Thus the picture suggests particles bearing charges could be modeled as defects
of vector fields.

The fact that charge--like conservation follows from a simple topological
construct, which depends only on continuity and dimension and pointing inside,
suggests that the construct has physical content.

It is this mathematical behavior which inspires us to suggest that a
physical
vector field whose index is not defined is involved with radiation. Also
we
suggest that the index is invariant under all changes of coordinates.
This
follows from a mathematical theorem.
\proclaim{theorem} Suppose $V$ is a space--like vector field in a
space--time
$M$. Suppose  $S$ and $T$ are two time--like slices of $M$ which can be
smoothly
deformed into each other. Suppose $D$, the defects of $V$, is compact in
the
region of $M$ where the deformation takes place. Then the index of $V$
projected
onto $S$ is equal to the index of $V$ projected onto $T$.
\endproclaim

This theorem is true since we can set up a proper otopy between the two
vector
fields given the hypotheses of the theorem. This means for any
space--like
vector field, the index is invariant under any choice of space--like
slices.
Thus it is mathematically true that the index is an invariant of general
relativity, just like proper time, unless there is topological radiation
or
there is a singularity or strange topology between the two slices so
that a
single coordinate system cannot represent the situation.

The last feature of the Principle of Invariance of Index to consider is
the
statement that the index should be invariant under a different choice of
orientation of space--time. This is not a mathematical consequence of
anything.
It just seems like a good guess. The orientation of space--time is a
man made choice. It shouldn't be involved with the underlying physics.
It is a broad statement and if it is
incorrect
it should be quickly apparent. At any rate it does predict the absence
of
magnetic monopoles.

\item{1.}Suppose we have a time--like vector field $T$ in a space--time
$M$.
Then the covariant derivative is orthogonal to $T$.
The zeros of this space--like field appear in all space--like slices
with the
same index. Vector fields can arise as gradients of functions, as duals
to
one--forms, as the dual of a two form contracted with a vector field, as
the
dual of a two--form operated on by $*d*$ where $*$ is the Hodge dual and
$d$ is the
exterior derivative. The $*$ of a three form is a one--form, but it is
one that
depends upon the choice of the orientation. The Hodge dual $*$ depends
upon
orientation, but any operation which involves an even number of Hodge
duals is
independent of orientation. The Principle of the Invariance of Index
then
implies that those physical vector fields which arise from one--forms
given by
formulas containing an odd number of Hodge dual operators should have
index zero
or undefined.

\item{2.} Apply the principle of Invariance of Index on vector fields
arising on
state spaces and configuration spaces.
We suggest studying pullbacks of equivariant vector fields on the
division algebras which look like Euclidean space.
Thus on the line, the complex plane, the quaternions, and the Cayley
numbers
there are equivariant vector fields corresponding to addition as well as
multiplication on the space minus the origin.
We should look at the pullbacks on a space--time pulled back by certain
functions from space--time to the division algebra.
Hopefully the symmetries of physics due to these Lie groups will be
reflected
in the properties of the pullback vector fields.
On the simplest level, if we have space $M$ and a potential function
$\phi$,
then the pullback of the constant vector field on the Real Numbers is
the
gradient of $\phi$ which is the force field.
There is another natural vector field on the Real Numbers which is an
equivariant vector field for the group of multiplications on $R-0$.
There is a function $v$ whose pullback of this equivariant vector field
is equal
to the gradient of $\phi$.
This $v$ turns out to be the speed of a particle obeying the potential
$\phi$.
\item{3.}We propose that familiar vector fields of physics should be
studied
and the motion of zeros and singularities be noted.
For example, the electrostatic field.
Given $E$,
we assign an integer to each point in space by introducing a constant
vector
field which creates a zero at that point.
The index of that zero is an integer.
The membrane between the positive integers and the negative integers
appears
interesting.
The relation of the zeros and the charges at equilibrium is
interesting.
We suggest that zeros closer to charges push the charges away, whereas
charges
closer to charges push the zeros away.
A very nice mathematical question in this regard is the following.  Let
$E$
be the electrostatic vector field of point charges satisfying Coulomb's
law.  Is there a configuration which will produce a zero  of index equal
to
any specified integer?  We can find zeros of index $1,-1$, and $0$, so
far, \cite{K}.

\item{4.} For the easiest cases of eigenfunction solutions to
Schrodinger's equation, it looks as if the gradient vector field of the
wave function for each energy level have different indices. If that is
so, then every change from one eigenstate to another results in a zero
or discontinuity flying out of or into infinity.

\item{5.} Suppose we have a fibre bundle and a proper vector field
defined on
the total space so that every vector is tangent along the fibre. What
kind of
vector fields on a given fibre could be the restriction of the global
vector
field along the fibre? In \cite{B-G} we show that there is a transfer on
homology whose trace is the index of the vector field restricted to the
fibre.
Thus the homology of the fibre bundle restricts the possible indices of
vector
fields on the fibre. For example, consider the principal $SU(2)$ --
bundle whose
total space is the 7 dimensional sphere and whose base space is the 4
dimensional sphere. This is the Hopf fibration. The homology only
permits
transfers of trace 0. Hence the index of the restriction of any vector
field
along the fibre restricted to a fibre must be zero.

Now vector fields along the fibre of a fibre bundle are a generalization
of the
concept of otopy. It is as if it represents a collection of possible
otopies under
certain circumstances. We say that a certain collection of zeros {\it
potentially interact}
on a fibre if they are contained in a connected component of zeros in
the total
space. Then the fact that a vector field along the fibre
restricts to a vector field of index zero
means
that the interacting zeros must have total index zero. Could some
argument such
as this be made to imply that there must be total charge zero in the
Universe?

\vfill\eject
\vfill\eject
\enddocument